# Presolar Silicon Carbide Grains of Types Y and Z: Their Strontium and Barium Isotopic Compositions and Stellar Origins

Nan Liu[1,2,3*], Thomas Stephan[4,5], Sergio Cristallo[6,7], Diego Vescovi[8], Roberto Gallino[9], Larry R. Nittler[3], Conel M. O'D. Alexander[3], Andrew M. Davis[4,5,10]

[1]Laboratory for Space Sciences and Physics Department, Washington University in St. Louis, St. Louis, MO 63130, USA; nliu@physics.wustl.edu

[2]McDonnell Center for the Space Sciences, Washington University in St. Louis, St. Louis, MO 63130, USA

[3]Earth and Planets Laboratory, Carnegie Institution for Science, Washington, DC 20015, USA

[4]Department of the Geophysical Sciences, The University of Chicago, Chicago, IL 60637, USA

[5]Chicago Center for Cosmochemistry, Chicago, IL, USA

[6]INAF-Osservatorio Astronomico d'Abruzzo, 64100 Teramo, Italy

[7]INFN-Sezione di Perugia, 06123 Perugia, Italy

[8]Institute for Applied Physics, Goethe University Frankfurt, 60438 Frankfurt, Germany

[9]Dipartimento di Fisica, Università di Torino, 10125 Torino, Italy

[10]The Enrico Fermi Institute, The University of Chicago, Chicago, IL 60637, USA

## ABSTRACT

We report the Sr and Ba isotopic compositions of 18 presolar SiC grains of types Y (11) and Z (7), rare types commonly argued to have formed in lower-than-solar metallicity asymptotic giant branch (AGB) stars. We find that the Y and Z grains show higher $^{88}$Sr/$^{87}$Sr and more variable $^{138}$Ba/$^{136}$Ba ratios than mainstream (MS) grains. According to FRANEC Torino AGB models, the Si, Sr, and Ba isotopic compositions of our Y and Z grains can be consistently explained if the grains came from low mass AGB stars with $0.15\ Z_\odot \leq Z < 1.00\ Z_\odot$, in which the $^{13}$C neutron exposure for the slow neutron-capture process is greatly reduced with respect to that required by MS grains for a 1.0 $Z_\odot$ AGB star. This scenario is in line with the previous finding based on Ti isotopes, but it fails to explain the indistinguishable Mo isotopic compositions of MS, Y, and Z grains.

## 1. INTRODUCTION

Spectroscopic and photometric data reveal the ubiquitous presence of silicon carbide (SiC) in the circumstellar envelopes of C-stars based on the 11.3 μm emission feature in their infrared spectra [1-3]. These SiC stardust grains contribute to the dust reservoir in the interstellar medium (ISM) and become part of the initial building blocks of stars forming in dense ISM regions if they survive destructive processes in the ISM. When it formed, the solar system incorporated such ancient stardust grains, which are preserved in small solar system bodies, e.g., primitive asteroids, that have not experienced significant internal heating since their formation. In primitive extraterrestrial materials from small solar system bodies, SiC and other types of stardust grains (e.g., silicates, oxides, graphite) are identified by their exotic isotopic compositions that cannot be explained by any chemical or physical processes occurring in the solar system and require origins around ancient stars (see [4, 5] for reviews). Since such stardust grains formed before the formation of the solar system in stellar winds or the debris of stellar explosions, they are commonly known as presolar grains. As bona fide stellar materials, presolar grains allow for detailed isotope analyses using modern mass spectrometric techniques in the laboratory that have become an important component of nuclear astrophysics [6].

Among various types of presolar phases, SiC is the most extensively studied. Thousands of presolar SiC grains have been examined for their C, N, and Si isotopic compositions, based on which the grains have been divided into five main groups, including mainstream (MS), Y, Z, AB, and X [4]. It is generally recognized that MS grains – the dominant group of presolar SiC grains (>~85% in number) – came from low-mass C-rich asymptotic giant branch (AGB) stars given (*i*) the slow neutron capture process (*s*-process) isotopic signatures commonly found in MS grains [7], (*ii*) the ubiquitous presence of SiC around such stars [1-3], and (*iii*) the similar ranges of C and N isotope ratios observed for MS grains and C-rich AGB stars [8]. In addition, X grains (1–2%) are thought to have come from core-collapse Type II supernovae based on the inferred incorporation of many short-lived nuclides, e.g., $^{44}$Ti [9, 10].

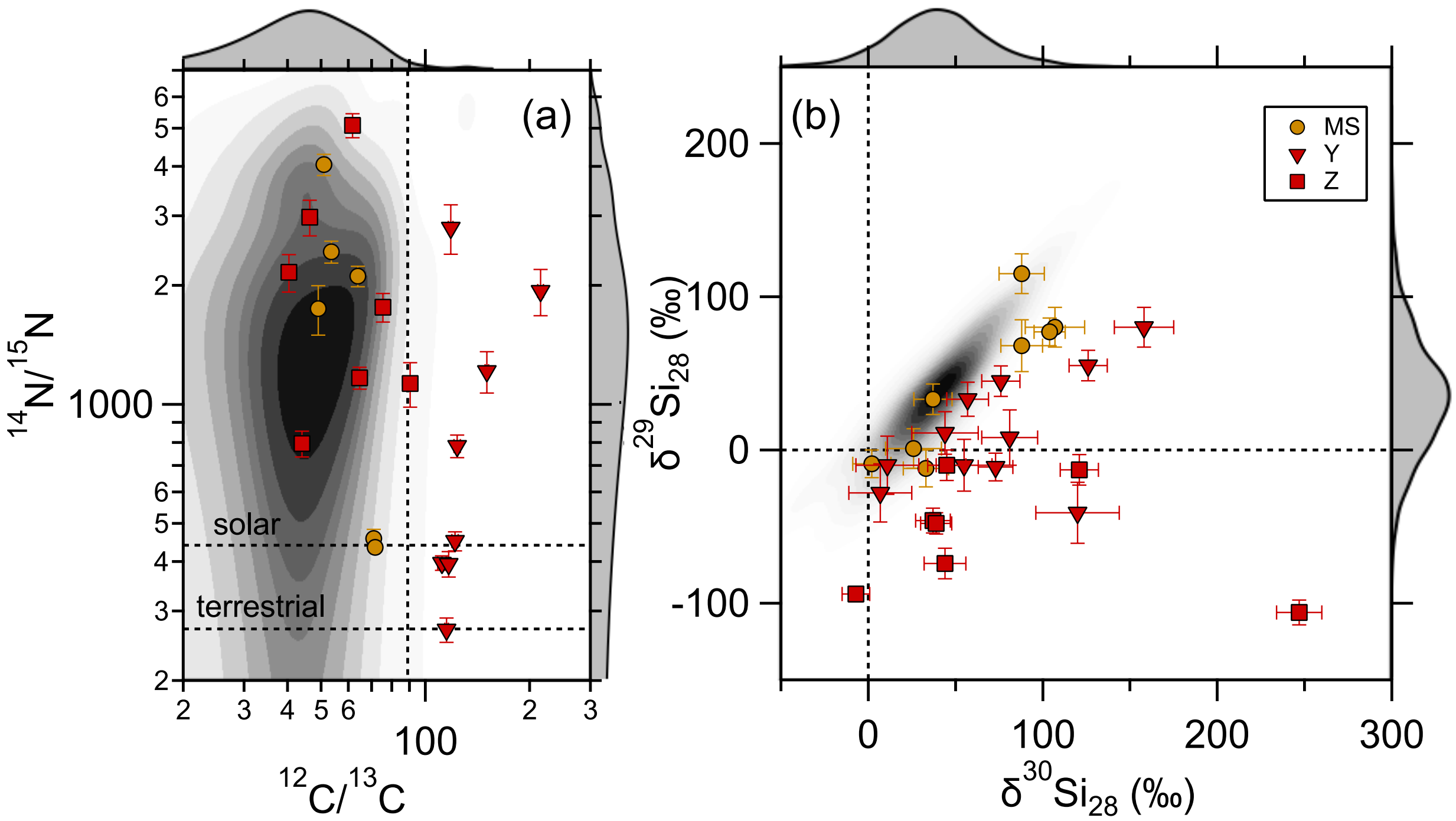

***Figure 1.*** *The C, N, and Si isotope ratios of MS, Y, and Z grains. The Y and Z grains from this study are compared to MS grains from [11]. In panel (a), the least contaminated MS grain data (for which N contamination was greatly suppressed) from [8] are shown as a greyscale density map (linearly increasing darkness with increasing density). In panel (b), high-precision MS grain data (1σ errors ≤ 10 ‰) from the Presolar Grain Database [12] are shown as a greyscale density map. Unless noted otherwise, the dashed lines represent the terrestrial composition. Errors are 1σ. All density maps in the figures of this study were generated by using the seaborn (version 0.11.2) jointplot function in Python (with default parameter values) based on a Gaussian kernel density estimator[1]. The grey histograms in all figures represent the respective calculated density distributions for MS grains.*

The stellar origins of AB (~5%), Y (1–6%), and Z (1–8%) grains are quite ambiguous, resulting from the lack of distinctive isotopic signatures and multielement isotope data (especially for heavy elements) [13, 14]. Recent studies [11, 15-20] suggest that AB grains – characterized by large $^{13}C$ excesses ($^{12}C/^{13}C \lesssim 10$) – consist of grains from J-type C-stars and core-collapse Type II supernovae and, possibly, born-again AGB stars. Type Y grains are defined to have $^{12}C/^{13}C \geq 100$, and type Z grains deviate from the MS grain line toward larger $^{30}Si$ excesses in a Si three-isotope plot (Fig. 1b). Previous studies suggest that the abundances of Y and Z grains increase with decreasing grain size [21]. Recent statistical analyses based on cluster analysis techniques [18, 19] pointed out that the classifications of MS, Y, and Z grains are somewhat arbitrary and not statistically significant. Based on Si and Ti isotopes, it has been long argued that types Y and Z grains came from low-mass AGB stars with initial metallicities that were lower (~1/3–1/2 $Z_\odot$) than those of MS grains (~$Z_\odot$) [13, 21, 22]. The proposed low-metallicity origins of types Y and Z grains, however, were recently challenged by the observation that their Mo isotopic compositions are indistinguishable from those of MS grains [23], in contrast to varying Mo isotopic patterns predicted by nucleosynthesis models for AGB stars with different metallicities. Here, we report Sr and Ba isotope data for Y and Z grains to provide the first piece of evidence that these two uncommon grain types show heavy-element isotopic compositions that are different from MS grains, namely higher $^{88}$Sr/$^{87}$Sr and more variable $^{138}$Ba/$^{136}$Ba ratios observed for Y and Z grains.

[1] See https://seaborn.pydata.org/generated/seaborn.kdeplot.html for details regarding the default value chosen for standard deviation of the smoothing kernel for the Gaussian kernel density estimator (kde) in seaborn kde plot.

## 2. AGB STELLAR NUCLEOSYNTHESIS MODELS

In this study, we will adopt two sets of AGB stellar models for comparison with our presolar SiC grain data, namely the magnetic FRUITY[2] AGB models presented in [27, 28] and the FRANEC Torino AGB models in [21, 23]. The magnetic FRUITY models are chosen for comparison with the Y and Z grain data from this study, because these models provide a good match to the heavy-element isotopic compositions of MS grains [27]. The magnetic FRUITY AGB models differ from the respective nonmagnetic FRUITY AGB models mainly in the physical model adopted for the $^{13}$C formation process. While nonmagnetic FRUITY AGB models consider convective overshooting to be the mechanism for driving the partial mixing of H into the He-intershell to form $^{13}$C [26], magnetic buoyancy is responsible for this process in magnetic FRUITY AGB models (see below for discussion in more detail). We also choose the FRANEC Torino AGB models for data-model comparisons because these models were used in the first systematic study of Y and Z grains for light-element isotopes [21], which led to the conclusion that Y and Z grains came from ~1/2 $Z_\odot$ and ~1/3 $Z_\odot$ AGB stars, respectively. The FRANEC Torino AGB models, which adopted updated solar system abundances [29] and nuclear reaction rates, were also used for comparison with the Mo isotopic compositions of Y and Z grains from our previous study [23].

Below, we provide a brief description of *s*-process nucleosynthesis in AGB stars. For detailed descriptions of *s*-process nucleosynthesis in AGB stars and associated modeling uncertainties, the reader is referred to [30, 31]. Stellar models have shown that the *s*-process operates in the He-intershell of low-mass (~1.5 $M_\odot \leq M \leq 3\text{–}4\ M_\odot$) AGB stars [32]. During the interpulse phase, the *s*-process is powered by the $^{13}C(\alpha,n)^{16}O$ reaction − the major neutron source for the *s*-process − at a neutron density of ~$10^7$–$10^8$ cm$^{-3}$ on a timescale of 5–20 ka. As shell H-burning proceeds, the He-intershell is heated and compressed, leading to the development of a thermal pulse (TP) when the temperature and density are high enough. During a TP, *s*-process products are further modified by neutron capture that is powered by the partial activation of the $^{22}Ne(\alpha,n)^{25}Mg$ reaction − the minor neutron source for the *s*-process − in the He-intershell, providing neutrons at a density of $10^9$–$10^{10}$ cm$^{-3}$ on a timescale of a few years. The short, high-density neutron exposure controls the production of nuclides affected by *s*-process branch points, at which neutron capture competes with beta decay due to their comparable rates [30, 33]. Despite the abovementioned consensus among AGB models, uncertainties in nuclear reaction rates and parameters in AGB stellar models lead to uncertain model predictions for the *s*-process. In particular, the formation of $^{13}C$, the major neutron source for the *s*-process, in the He-intershell is a fundamental unknown that is directly related to the *s*-process nucleosynthesis (see [28, 31, 34-36] for discussion). Below, we focus on discussing differences between the two sets of AGB models.

[2] FRUITY is based on FRANEC (Frascati Raphson-Newton Evolutionary Code) code [24] and stands for FRANEC Repository of Updated Isotopic Tables & Yields [25, 26]. The nonmagnetic FRUITY models are available online at http://fruity.oa-teramo.inaf.it/, while the magnetic FRUITY models are not available online yet.

First, while the FRANEC Torino AGB models were based on the FRANEC stellar code [24, 37], the magnetic FRUITY AGB models were based on the FUNS[3] stellar code, which differs from FRANEC in several important details. In particular, FUNS adopted molecular opacities that consider increasing opacity with the formation of C- and N-bearing molecules at low temperatures [39]. Also, the adopted mass loss rate in FUNS was calibrated against the physical properties of a sample of Galactic giant stars [38] and differs from the mass loss law adopted in FRANEC. In addition, FRUITY stellar models were computed by coupling a full nuclear network to the FUNS stellar evolution code [25, 26], in contrast to the postprocessing approach adopted in the Torino models [32]. In comparison to the FRANEC Torino stellar models, the FRUITY stellar models are characterized by higher third dredge-up (TDU) efficiencies, higher mass loss rates, and, in turn, lowered maximum stellar temperature ($T_{max}$) in the He-intershell.

[3] FUNS stands for FUll Network Stellar and is a more recent version of the original FRANEC code [24]. For a full description of the FUNS code, we refer the reader to [38].

Second, the two sets of AGB stellar models also adopted slightly different nuclear reaction rates. The FRANEC Torino models in this study adopted ($n$,$\gamma$) cross sections that were recommended by KADoNiS v0.3[4], while the FRUITY models adopted ($n$,$\gamma$) cross sections mainly recommended in [40] with recent updates compiled in [41]. Regarding Sr and Ba isotopes, the two sets of ($n$,$\gamma$) cross sections are essentially the same. In addition, while the FRANEC Torino models adopted the $^{22}$Ne($\alpha$,$n$)$^{25}$Mg and $^{22}$Ne($\alpha$,$\gamma$)$^{26}$Mg rates recommended by [42] and [43], respectively, the magnetic FRUITY models adopted those recently recommended by [44], which are lower than the former (*e.g.*, by a factor of 2.5 at $3 \times 10^8$ K for $^{22}$Ne($\alpha$,$n$)$^{25}$Mg). The lowered $^{22}$Ne($\alpha$,$n$)$^{25}$Mg rate adopted in the FRUITY stellar models, together with the lowered $T_{\rm max}$, results in a much less efficient operation of the $^{22}$Ne($\alpha$,$n$)$^{25}$Mg reaction during TPs in the magnetic FRUITY stellar models than in the corresponding FRANEC Torino models.

Third, the two sets of AGB models adopted different formulae for partial mixing of H from the envelope into the He-intershell, which allows for the formation of $^{13}$C via the $^{12}$C($p$,$\gamma$)$^{13}$N($\beta^+$)$^{13}$C reaction chain. Carbon-13 is the major neutron source for the *s*-process. The upper thin layer of the He-intershell that contains $^{13}$C, is often referred to as the "$^{13}$C pocket" (a few $10^{-3}$ $M_\odot$ in mass). The magnetic FRUITY models adopted the formula for magnetic-buoyancy-driven mixing presented in [35, 45] to allow a partial mixing of H into the He-intershell (see [27] for details). In all the magnetic FRUITY models presented here, the values for two parameters – magnetic strength $B_\varphi$ and velocity of uprising magnetic flux tubes $u_{\rm p}$ – were fixed at $5 \times 10^4$ G and $5 \times 10^{-5}$ cm/s, respectively, which were calibrated against the heavy-element isotopic compositions of MS grains [27]. Since the FRUITY stellar models were computed in a fully coupled way, the $^{13}$C pocket profile varied as the AGB star evolved. In the FRANEC Torino AGB models, the H mixing velocity was assumed to follow an exponentially decaying profile as a result of convective overshooting. Convective overshooting leads to a partial mixing of H into the He-intershell as convective eddies cross the bottom of the convective envelope and move downward into the He-intershell with an exponentially decaying velocity [26, 38, 46-48]. The $^{13}$C pocket was implemented by using a three-zone scheme with a total $^{13}$C pocket mass of $1 \times 10^{-3}$ $M_\odot$ (see [32] for details). The mixed-in H concentration in the He-intershell was considered as a free parameter and was simultaneously increased or decreased in the three zones by different factors, corresponding to different cases [32, 49]. The ST (for "standard") case was so named for historical reasons [50] and used as the reference case. Since the FRANEC Torino models adopted a postprocessing approach, the $^{13}C$ pocket was unchanging during the AGB phase. As shown in [27], the formula for the mixing velocity adopted in the magnetic FRUITY models led to the formation of a $^{13}C$ pocket with a larger mass (~$3\times10^{-3}$ $M_\odot$) and a power-law dependence of the $^{13}C$ concentration as a function of stellar radius, in contrast to the exponential decaying $^{13}C$ profile based on convective overshooting.

[4] KADoNiS stands for Karlsruhe Astrophysical Database of Nucleosynthesis in Stars. Version 0.3 is available at https://www.kadonis.org/ and version 1.0 is available at https://exp-astro.de/kadonis1.0/.

## 3. RESULTS

The presolar SiC grains analyzed in this study were separated from the CM2 chondrite Murchison using the CsF dissolution method described by [51]. We identified a total of 33 Y grains and 28 Z grains by imaging thousands of grains for C, N, and Si isotopes with the Cameca NanoSIMS 50L instrument at the Carnegie Institution. Subsequently, using the Chicago Instrument for Laser Ionization (CHILI) [52], we obtained sufficiently precise Sr and Ba isotopic compositions for 11 of the Y grains and seven of the Z grains. The Sr and Ba data reported here (Table 1) were obtained in the same analytical session as those of 19 Y, 18 Z, 16 AB1, 12 AB2, and 15 MS grains reported by [11, 16, 23]. The details for sample preparation and CHILI analyses were given by [11] and [53], respectively. For 12 of the 18 Y and Z grains in Table 1, their correlated Mo isotopic compositions can be found in [23]. The Si, Sr, and Ba isotope ratios are reported in delta notation defined as $\delta^{i}A_{j}$ (‰) = $[(^{i}A/^{j}A)_{grain} / (^{i}A/^{j}A)_{std} - 1] \times 1000$, in which $(^{i}A/^{j}A)_{grain}$ and $(^{i}A/^{j}A)_{std}$ denote the measured isotope ratios for a grain and a standard, respectively. For Si, Sr, and Ba, the denominator isotopes are $^{28}Si$, $^{87}Sr$, and $^{136}Ba$, respectively. The choice of $^{87}Sr$ for calculating Sr isotope ratios in this study differs from the common use of $^{86}Sr$ in previous studies [53-55], and was done because we observed an unidentified molecular interference peak at mass 86 u in the mass spectra of some grains. In addition, since we first conducted NanoSIMS analyses using a $Cs^+$ beam and implanted $^{133}Cs^+$ into the SiC grains from this study, we noticed a few counts at mass 133 u but no interference at masses 134 u and 135 u in any of the CHILI mass spectra. Values of $\delta^{130,132,134}Ba_{136}$ are not reported in Table 1 because of large statistical uncertainties that are caused by the low abundances of these isotopes. All the isotope data are reported with 1σ errors in Table 1.

**Table 1.** Isotope Data of Y and Z Grains[a,b].

| Grain | Group | Size ($\mu m^2$) | $^{12}C/^{13}C$ | $^{14}N/^{15}N$ | $\delta^{29}Si_{28}$ (‰) | $\delta^{30}Si_{28}$ (‰) | $\delta^{84}Sr_{87}$ (‰) | $\delta^{88}Sr_{87}$ (‰) | $\delta^{135}Ba_{136}$ (‰) | $\delta^{137}Ba_{136}$ (‰) | $\delta^{138}Ba_{136}$ (‰) |
|---|---|---|---|---|---|---|---|---|---|---|---|
| M1-A4-G473 | Y | 1.5×1.3 | 118.6±1.8 | 2795±398 | −28±29 | 7±18 | — | — | −811±133 | −304±220 | −150±201 |
| M1-A5-G693 | Y | 1.0×0.7 | 115.2±1.3 | 269±19 | −10±19 | 11±18 | — | — | −624±146 | −371±175 | −152±174 |
| M1-A5-G879 | Y | 0.7×1.0 | 215.4±3.9 | 1935±261 | −41±20 | 120±24 | −989±34 | 101±38 | −834±12 | −563±18 | −422±16 |
| M1-A5-G1096 | Y | 0.7×1.2 | 117.0±1.1 | 394±29 | 8±18 | 81±16 | — | — | −880±98 | −504±165 | −317±160 |
| M1-A7-G812 | Y | 0.8×0.9 | 150.7±1.8 | 1213±145 | −10±17 | 55±16 | — | — | −757±101 | −581±100 | −364±102 |
| M2-A1-G176 | Y | 1.0×1.6 | 111.7±3.2 | 397±17 | 45±10 | 76±11 | −887±107 | 154±61 | −872±38 | −548±67 | −33±95 |
| M2-A2-G262 | Y | 1.2×0.8 | 148.4±3.3 | — | 80±13 | 158±17 | −354±187 | 123±56 | −687±101 | 57±203 | 346±190 |
| M2-A2-G644 | Y | 0.8×1.0 | 130.2±6.0 | — | 11±14 | 44±19 | −819±105 | –62±38 | −766±129 | −547±134 | −318±115 |
| M2-A2-G1140[c] | Y | 1.2×1.2 | 120.3±1.3 | — | 33±11 | 57±12 | — | — | *−142±90* | *−8±84* | *−67±65* |
| M3-G281 | Y | 1.2×1.3 | 123.6±2.7 | 784±51 | −11±9 | 73±10 | — | — | −759±144 | −642±166 | −221±211 |
| M3-G1207 | Y | 1.3×1.5 | 121.8±2.6 | 451±25 | 55±10 | 126±11 | — | — | −420±197 | −613±126 | 166±247 |
| M2-A1-G469 | Z | 0.7×0.7 | 64.7±1.8 | 1166±75 | −106±8 | 247±13 | — | — | — | −882±155 | −700±150 |
| M2-A2-G791 | Z | 0.7×0.7 | 46.4±1.0 | 2974±306 | −46±8 | 37±10 | −847±117 | 198±62 | −721±63 | −556±91 | −202±84 |
| M2-A4-G1220 | Z | 0.8×0.9 | 44.1±1.0 | 795±60 | −48±7 | 39±9 | — | — | −732±80 | −474±91 | −305±83 |
| M3-GB4 | Z | 0.7×0.7 | 75.4±1.6 | 1759±146 | −74±10 | 44±12 | — | — | −736±82 | −422±111 | 140±152 |
| M3-G628 | Z | 1.8×2.0 | 40.4±0.2 | 2156±233 | −94±5 | −7±8 | — | — | −860±85 | −323±151 | 97±180 |
| M3-G692 | Z | 0.7×0.6 | 90.4±2.1 | 1128±146 | −10±10 | 45±11 | −659±223 | –9±92 | −624±149 | −500±146 | −361±134 |
| M3-G1519 | Z | 0.8×0.6 | 61.6±0.4 | 5073±351 | −13±11 | 184±28 | −785±144 | 129±89 | −932±94 | −422±134 | −79±154 |

Note: [a]For 12 of the 18 Y and Z grains in the table, their C, N, Si, and Mo isotope data were previously reported in [23].

[b]All data are reported with 1σ errors.

[c]The Mo isotopic composition of grain M2-A2-G1140 suggests that its Ba isotope data are dominated by Ba contamination (see Appendix for discussion in detail). Thus, the Ba isotope data of grain M2-A2-G1140 is excluded in figures for comparison with AGB models.

Figure 1a shows that a higher percentage of our Y grains exhibits $^{14}N/^{15}N$ ratios between terrestrial and solar values when compared to our Z grains and the literature MS grains from [8]. The literature MS grain data from [6] were obtained after an extended period of ion sputtering, which was shown to effectively reduce sampling surface N contamination. Since we did not adopt such an analytical protocol during the N isotope analyses of the Y and Z grains from this study and the MS grains from [11], the terrestrial-to-solar $^{14}N/^{15}N$ ratios observed in four of our eight Y grains in Fig. 1a could have been caused by sampling significant asteroidal and/or terrestrial N contamination during the Y grain analyses. It remains to be seen whether uncontaminated MS, Y, and Z grains show any differences in $^{14}N/^{15}N$. Currently, there is a lack of a quantitative and consistent definition for Z grains [12]. The Z grains from this study are characterized by lower-than-terrestrial $^{29}Si/^{28}Si$ ratios and >3.5σ deviations from the MS trend (greyscale map in Fig. 1b).

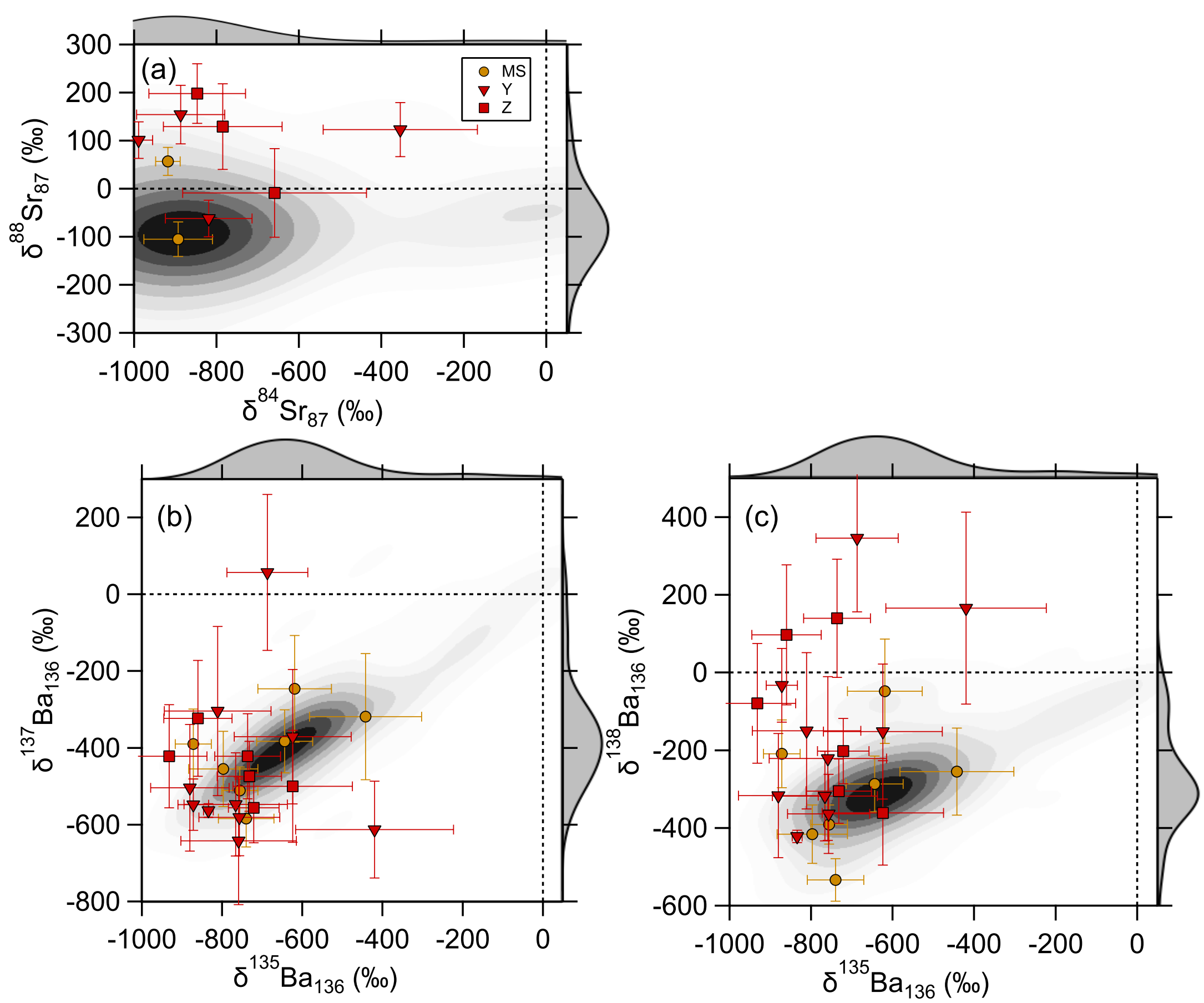

***Figure 2.*** *The Sr and Ba isotopic compositions of MS, Y, and Z grains. The Y and Z grains from this study are compared to MS grains from [11], all of which were analyzed in the same CHILI session. MS grains from other previous studies [34, 53, 54, 56] are shown as greyscale density maps. Errors are 1σ.*

Figure 2 compares the Sr and Ba isotopic compositions of the 18 Y and Z grains from this study to those of the ten MS grains from [11] (one MS grain with errors >200‰ is not included). Figure 2 clearly shows that, like MS grains, Y and Z grains carry *s*-process Sr and Ba isotopic signatures, further corroborating their AGB stellar origins. Furthermore, we see in Fig. 2 that almost all grains show $\delta^{84}Sr_{87} < -600‰$ and $\delta^{135}Ba_{136} < -400‰$, which is in line with the literature data for acid-cleaned [34, 53, 54] and "uncontaminated" MS grains (see [56] for details) and thus implies no substantial amounts of terrestrial/asteroidal contamination for Sr or Ba sampled during our analyses (except for the Ba isotope data of Y grain M2-A2-1140; see Fig. A1 and discussion in Appendix). Finally, while we observe no significant differences in $\delta^{84}Sr_{87}$ and $\delta^{135,137}Ba_{136}$ between Y/Z and MS grains (Fig. 2), our Y and Z grains overall exhibit higher $\delta^{88}Sr_{87}$ values and more variable $\delta^{138}Ba_{136}$ values when compared to the literature MS grain data (greyscale density maps in Fig. 2).

## 4. DISCUSSION

### *4.1. s-Process Production of Sr and Ba Isotopes*

The reader is referred to [34, 54] for detailed discussions of the *s*-process production of Sr and Ba isotopes in AGB stars. Here, we provide a brief overview focusing on the effects of bottleneck isotopes along the *s*-process path and branch points where neutron capture competes with $\beta^-$ decay in Sr and Ba mass regions. For discussions of light-element isotope productions in AGB stars, the reader is referred to [21, 57].

Strontium has four stable isotopes: $^{84}$Sr, $^{86}$Sr, $^{87}$Sr, and $^{88}$Sr. The proton-rich isotope $^{84}$Sr is shielded from the *s*-process path, and the low $^{84}$Sr/$^{87}$Sr ratios of AGB SiC grains in Fig. 2 are caused mainly by the overproduction of $^{87}$Sr, which is a pure *s*-process isotope. Although the radioactive nuclide $^{87}$Rb decays to $^{87}$Sr with a half-life of 49.2 Ga, we do not expect any noticeable radiogenic contribution from $^{87}$Rb to $^{87}$Sr in presolar SiC grains since the volatility of Rb is similar to that of Cs, which is absent in presolar SiC grains [7, 34] (see section 5.4.3 in [58] for discussion in detail). The neutron-rich isotope $^{88}$Sr has a magic number of neutrons ($N$ = 50) and thus a particularly stable nuclear structure, resulting in a small neutron capture cross section (15 times smaller than that of the adjacent isotope $^{87}$Sr). There is an important branch point in the Kr-Rb-Sr region at $^{85}$Kr (isomeric state, $t_{1/2}$ = 4.5 hours; ground state, $t_{1/2}$ = 11 years). This results in two main $s$-process channels in this region: (1) $^{85}$Kr($\beta^-\bar{\nu}_e$)$^{85}$Rb($n,\gamma$)$^{86}$Rb($\beta^-\bar{\nu}_e$)$^{86}$Sr($n,\gamma$)$^{87}$Sr($n,\gamma$)$^{88}$Sr and (2) $^{85}$Kr($n,\gamma$)$^{86}$Kr($n,\gamma$) $^{87}$Kr($\beta^-\bar{\nu}_e$)$^{87}$Rb($n,\gamma$)$^{88}$Rb($\beta^-\bar{\nu}_e$)$^{88}$Sr.

Barium has seven stable isotopes: $^{130}$Ba, $^{132}$Ba, $^{134}$Ba, $^{135}$Ba, $^{136}$Ba, $^{137}$Ba, and $^{138}$Ba. The proton-rich isotopes $^{130}$Ba and $^{132}$Ba are shielded from the $s$-process path. Barium-134 and $^{136}$Ba both are pure $s$-process isotopes, but the $^{134}$Ba/$^{136}$Ba ratio produced by AGB nucleosynthesis is affected by a branch point at $^{134}$Cs ($t_{1/2}$ = 2.1 years), whose $\beta^-$ decay rate is a strong function of temperature [59-61]. This results in two main $s$-process channels: (1) $^{133}$Cs($n,\gamma$)$^{134}$Cs($\beta^-\bar{\nu}_e$)$^{134}$Ba($n,\gamma$)$^{135}$Ba($n,\gamma$)$^{136}$Ba and (2) $^{133}$Cs($n,\gamma$)$^{134}$Cs($n,\gamma$)$^{135}$Cs($n,\gamma$) $^{136}$Cs($\beta^-\bar{\nu}_e$)$^{136}$Ba. Like $^{88}$Sr, the neutron-rich isotope $^{138}$Ba also has a magic number of neutrons ($N$ = 82) and thus a small neutron capture cross section.

The $s$-process theory [62, 63] predicts that the product of $\sigma_A N_A$, in which $\sigma_A$ is the Maxwellian-averaged neutron capture cross section of a nuclide with mass A and $N_A$ its $s$-process production, remains approximately constant during $s$-process nucleosynthesis, given the low neutron densities for the $s$-process ($10^7$–$10^8$ cm$^{-3}$). In turn, it predicts that the $s$-process production of nuclide A is inversely correlated with its $\sigma_A$ value. The few exceptions to this steady-state scenario for the $s$-process are isotopes with magic numbers of neutrons, which have small neutron capture cross sections and act as bottlenecks along the $s$-process. The most important bottlenecks, namely $^{88}$Sr (50 neutrons), $^{138}$Ba (82 neutrons), and $^{208}$Pb (82 protons and 126 neutrons; doubly magic), along the $s$-process cause accumulation of neutrons at mass 88 u, 138 u, and 208 u, resulting in the three $s$-process peaks for the solar system isotope abundances [64]. Given the bottleneck effects, the abundances of $^{88}$Sr and $^{138}$Ba are not simply inversely correlated with their respective $\sigma_A$ values [6]. It was shown that the relative $s$-process productions of $^{88}$Sr and $^{138}$Ba are sensitive to the detailed distribution of the major neutron source $^{13}$C in the $^{13}$C pocket [34, 54]. The activation strength of branch points, on the other hand, is dominantly controlled by the short, high-density neutron exposure released by the minor neutron source $^{22}$Ne($\alpha,n$)$^{25}$Mg during TPs [30, 33]. In summary, while the ratios of $^{88}$Sr/$^{87}$Sr and $^{134}$Ba/$^{136}$Ba are affected by branching effects, the ratios of $^{138}$Ba/$^{136}$Ba and $^{88}$Sr/$^{87}$Sr (additionally) are affected by the distribution of $^{13}$C in the $^{13}$C pocket.

In the following sections, we will compare the isotopic compositions of MS, Y, and Z grains to the two sets of stellar models for AGB stars with varying metallicities, namely the magnetic FRUITY AGB models and FRANEC Torino AGB models. Since magnetic FRUITY

AGB models, so far, have been run for 2 $M_{\odot}$ stars only [27, 28], we will conduct the data-model comparison focusing on 2 $M_{\odot}$ models to investigate the differences between the two sets of models. We will include an additional 3 $M_{\odot}$ FRANEC Torino AGB model to illustrate the effect of the initial stellar mass on AGB model predictions for Si, Ti, Sr, and Ba isotope ratios.

*4.2. Magnetic FRUITY AGB Models*

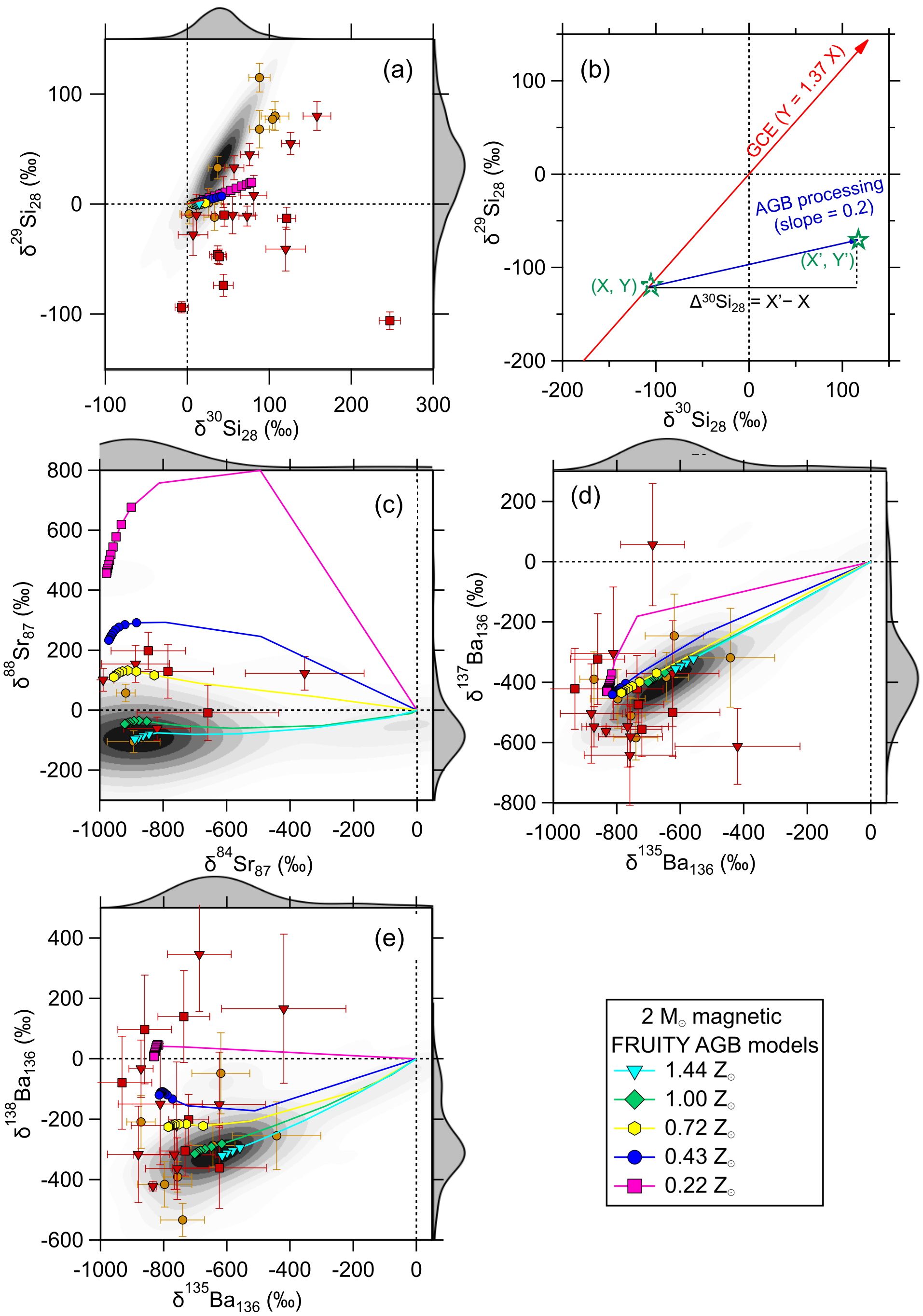

***Figure 3***. *In panels (a), (c), (d), and (e), three-isotope plots compare the same set of grain data as in Fig. 2 to magnetic FRUITY model calculations for Si, Sr, and Ba isotopes. For the models, lines represent O-rich phases and lines with symbols represent C-rich phases, during which SiC is expected to most likely condense [65]. Each symbol represents a TP. The solar metallicity (i.e., the mass fraction of elements heavier than He in the solar system) refers to 0.014 [66]. Errors are all 1σ. In panel (b), we illustrate that $\Delta^{30}Si_{28}$ is defined as the horizontal distance between the initial envelope composition of the parent AGB star of a grain along the GCE (Galactic chemical evolution) line and the final envelope composition of the star from which the grain condensed.*

The magnetic FRUITY models in Fig. 3 predict increasing $\delta^{88}Sr_{87}$ values with decreasing initial stellar metallicity, resulting from the increasing $^{13}$C/Fe ratio and thus increasing *s*-process efficiency with decreasing metallicity [49]. The predicted trend of decreasing $\delta^{88}Sr_{87}$ values with increasing TPs results from the combined effects of (*i*) convective burning of leftover $^{13}$C in the $^{13}$C pocket during the first one or two TPs and (*ii*) the shrinking of the $^{13}$C pocket with increasing TPs following the natural shrinking of the He-intershell region [25, 26]. Thus, in the magnetic FRUITY models, the first $^{13}$C pocket is the largest, and the amount of unburned $^{13}$C in the first $^{13}$C pocket leads to the highest neutron density during the first TP, the strongest activation of the path (2) $^{85}$Kr($n,\gamma$)$^{86}$Kr($n,\gamma$)$^{87}$Kr($\beta^- \bar{\nu}_e$)$^{87}$Rb($n,\gamma$)$^{88}$Rb($\beta^- \bar{\nu}_e$)$^{88}$Sr, and, in turn, the highest model prediction for $\delta^{88}Sr_{87}$ at the first TP. Subsequently, given the shrinking $^{13}$C pocket (i.e., decreasing amount of $^{13}$C) and the limited activation of the minor neutron source during TPs in the magnetic FRUITY stellar models, the model predictions for $\delta^{88}Sr_{87}$ gradually decrease with increasing TPs. This effect becomes more evident in low-metallicity models because the neutron-to-seed ratio (i.e., $^{13}$C/Fe) for the *s*-process increases linearly with decreasing metallicity.

Figure 3 shows that Y and Z grains have higher $\delta^{88}Sr_{87}$ values than MS grains, implying lower-metallicity stellar origins of the Y and Z grains by a factor of two on average. Specifically, Fig. 3c implies that the MS grains came from AGB stars with initial metallicities of 1.19−1.43 $Z_\odot$ and that the Y and Z grains came from AGB stars with $0.58\ Z_\odot \leq Z < 1.43\ Z_\odot$. A positive correlation between $\delta^{88}Sr_{87}$[5] and the initial metallicity is predicted by all existing AGB models and supported by the *s*-process element enrichments of barium stars [67]. The accuracy in the derived initial stellar metallicities for the MS, Y, and Z grains based on Fig. 3c, however, is affected by uncertainties in $T_{\max}$, nuclear reaction rates, and parameters in the physical model for the $^{13}$C pocket formation. This is because (*i*) model predictions for $\delta^{88}Sr_{87}$ are affected by the branch point at $^{85}$Kr and thus the efficiency of the minor neutron source $^{22}$Ne($\alpha,n$)$^{25}$Mg, which depends strongly on the $T_{\max}$ in the He-intershell during TPs, (*ii*) model predictions for $\delta^{88}Sr_{87}$ are directly affected by the $^{22}$Ne($\alpha,n$)$^{25}$Mg reaction rate and the neutron capture and $\beta^-$ decay rates of the reactions along the two main *s*-process paths for Sr isotopes (Section 4.1), and (*iii*) model predictions for $\delta^{88}Sr_{87}$ are also directly affected by the amount of $^{13}$C in the $^{13}$C pocket. In conclusion, although it is impossible to provide an accurate constraint on the initial metallicities of the parent stars of the MS, Y, and Z grains, the higher $\delta^{88}Sr_{87}$ values of the Y and Z grains suggest their origins in lower-metallicity AGB stars when compared to the MS grains.

[5] $\delta^{88}Sr_{86}$ values in [67] were calculated using $^{86}$Sr as the denominator isotope and differs from $\delta^{88}Sr_{87}$ values in this study. However, since $^{86}$Sr and $^{87}$Sr are both pure *s*-process isotopes and produced together along the same *s*-process path (see Section 4.1), $\delta^{88}Sr_{86}$ and $\delta^{88}Sr_{87}$ values are expected to show the same dependence on the initial stellar metallicity.

The magnetic FRUITY models for AGB stars with 0.58 $Z_{\odot} \leq Z < 1.43\ Z_{\odot}$, which explain the heavy-element isotopic compositions of the MS, Y, and Z grains in Fig. 3, however, cannot also explain the differences in Si isotopes between the Y/Z and MS grains. This point is better illustrated in Fig. 4, in which Sr and Ba isotope ratios are plotted against $\Delta^{30}Si_{28}$, which is a measure of $^{30}$Si excess produced by AGB stellar nucleosynthesis and represents the horizontal distance between the grain data point and the initial composition of its parent star on the Galactic chemical evolution (GCE) line (Fig. 3b; initially defined by [22]). The Si isotope ratios of the MS, Y, and Z grains receive contributions from both GCE and AGB stellar nucleosynthesis [51]. Here, we consider that the GCE evolves along a line with a slope of 1.37 and crosses the solar composition [68], and the FRUITY models predict that AGB stellar nucleosynthesis shifts the envelope composition away from the GCE trend toward $^{30}$Si excesses along a slope-0.2 line (Fig. 3b). Thus, $\Delta^{30}Si_{28}$ can be calculated from the equation $\Delta^{30}\mathrm{Si}_{28} = \frac{1.37\times\delta^{30}Si_{28} - \delta^{29}Si_{28}}{1.17}$. Although the slopes of both the GCE trend and the AGB evolution trend are subject to uncertainty and different choices of both could yield different $\Delta^{30}Si_{28}$ values, this should not affect the relative $\Delta^{30}Si_{28}$ differences between the MS, Y, and Z grains, which is the focus of our discussion here.

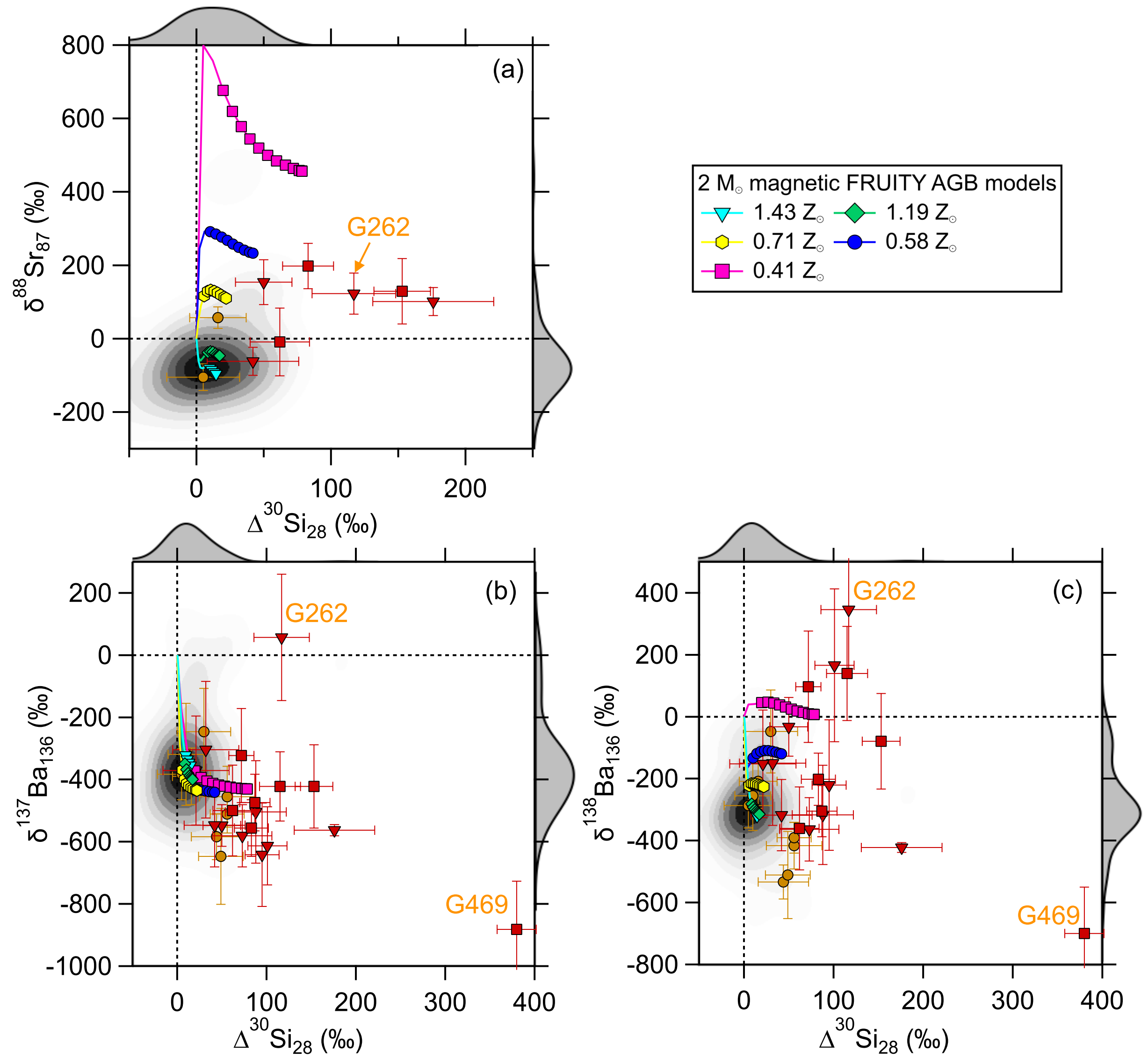

***Figure 4**. Plots of Sr and Ba isotope ratios versus $\Delta^{30}Si_{28}$ (see text for its definition) comparing the same sets of grain data with the same magnetic FRUITY AGB models as in Fig. 3. Y grain M2-A2-G262 and Z grain M2-A1-G469 are labled as G262 and G469, respectively.*

Figure 4 clearly shows that (*i*) none of the magnetic FRUITY models can explain the large $\Delta^{30}Si_{28}$ values of the Y and Z grains by AGB stellar nucleosynthesis, and (*ii*) although the difference in predicted $\delta^{138}Ba_{136}$ values between the 1.43 and 0.41 $Z_\odot$ models can account for the difference observed between the MS and Y/Z grains, the 0.41 $Z_\odot$ model predicts too high $\delta^{88}Sr_{87}$ to explain the Y and Z grain data. Previous studies [21, 23] showed that AGB model predictions for $\Delta^{30}Si_{28}$ are solely controlled by the efficiency of the minor neutron source $^{22}Ne(\alpha,n)^{25}Mg$ and are barely affected by the adopted $^{13}C$ pocket (Fig. 5). Since the Y and Z grains overall show higher $\Delta^{30}Si_{28}$ and $\delta^{88}Sr_{87}$ values than MS grains (Fig. 4a), this implies that the higher $\delta^{88}Sr_{87}$ values of the Y and Z grains result dominantly from the enhanced efficiency of the $^{22}Ne(\alpha,n)^{25}Mg$ reaction instead of from the enhanced $^{13}C/Fe$ ratio in their parent stars.

The data-model discrepancies in $\Delta^{30}Si_{28}$ for the Y and Z grains in Fig. 4 are unlikely to be caused by uncertainties in the $^{22}Ne(\alpha,n)^{25}Mg$ reaction rate. The magnetic FRUITY AGB models adopted the new $^{22}Ne(\alpha,n)^{25}Mg$ and $^{22}Ne(\alpha,\gamma)^{26}Mg$ reaction rates from [44] at relevant AGB temperatures. In comparison, the nonmagnetic FRUITY AGB models (available at FRUITY database) adopted the rates from [42] and [43], respectively, which are the same as those adopted in the FRANEC Torino AGB models (Fig. 5) and should, in principle, result in a more effective operation of the $^{22}Ne(\alpha,n)^{25}Mg$ reaction than in the magnetic FRUITY AGB models (Fig. 4). However, the predicted $\delta^{30}Si_{28}$ values by the two sets of FRUITY models differ by only up to 4 ‰, emphasizing that the $^{22}Ne(\alpha,n)^{25}Mg$ reaction barely operates during TPs due to the low $T_{max}$ values in the FRUITY stellar models. In the next section, we investigate whether uncertainties in stellar models can account for the discrepancies observed between the Y/Z grain data and the magnetic FRUITY models by adopting the FRANEC Torino AGB models for comparison.

### *4.3. FRANEC Torino AGB Models*

The FRANEC Torino models predict larger $^{30}Si$ excesses at the stellar surface than the FRUITY models, because, compared to FUNS stellar models, FRANEC stellar models can reach higher $T_{max}$ and experience more TPs (see Section 2 for details). Figure 5 reveals that the MS grains can be explained by 2 $M_\odot$, 1.0 $Z_\odot$ FRANEC Torino AGB model calculations in the D1.5 to U1.3 cases. In comparison, the Si, Sr, and Ba isotopic compositions of the Y and Z grains could be consistently explained if these two rare types came from $0.15\ Z_\odot \leq Z < 1.00\ Z_\odot$ AGB stars in which the amount of $^{13}C$ in the $^{13}C$ pocket is reduced by up to a factor of 7.8 (D6 to D1.5 cases) relative to that is required by the MS grain data for a 1.0 $Z_\odot$ AGB star (D1.5 to U1.3 cases). This observation is, in fact, in line with the previous finding [21] that the $^{49}Ti$ and $^{50}Ti$ excesses of Y and Z grains are more compatible with Torino AGB model calculations in the D6 case than in the ST case.

In Fig. 6, the Torino AGB models that provide a good match to the grain data in Fig. 5 are further compared to MS, Y, and Z grains from previous studies [13, 21, 69-74] for Ti isotopes. Like Si isotopes, the abundances of Ti isotopes in AGB stellar envelope are also significantly affected by the GCE. Thus, $\Delta^{50}Ti_{48}$ (like $\Delta^{30}Si_{28}$ in Fig. 4b) is defined to represent $^{50}Ti$ excess produced by the *s*-process in AGB stars after the effect of GCE is corrected. Here, we consider that (*i*) the AGB stellar nucleosynthesis follows a trend with a slope of 0.24 (Fig. 6a) according to the magnetic FRUITY AGB models that provide a good match to the heavy-element isotopic compositions of MS grains[6] (Figs. 3, 4) and (*ii*) the GCE trend has a slope of 0.65 (Fig. 6a), which is a rough estimate based on the MS grains that have the smallest $\delta^{50}Ti_{48}$ values (with respect to $\delta^{46}Ti_{48}$) in Fig. 6a. Given these prerequisites, $\Delta^{50}Ti_{48}$ values in Fig. 6b were calculated using the equation $\Delta^{50}Ti_{48} = \frac{0.65\times\delta^{50}\mathrm{Ti}_{48} - \delta^{46}\mathrm{Ti}_{48}}{0.41}$. Figure 6b demonstrates that the Torino AGB models that match the MS, Y, and Z grain data in Fig. 5 also provide a satisfactory explanation for the Ti isotopic compositions of the three groups of grains.

The absolute $\Delta^{50}Ti_{48}$ values of the Y and Z grains, however, are directly affected by uncertainties in the assumed GCE and AGB stellar nucelsoynthesis trends. Different from Si isotopes, AGB stellar nucleosynthesis is predicted to follow trends with varying slopes in Fig. 6a, depending on the initial stellar metallicity and stellar mass, and also the $^{13}$C/Fe ratio [21]. For instance, the 2 $M_\odot$ and 3 $M_\odot$ FRANEC Torino AGB models for a 0.3 $Z_\odot$ star in the D3 case predict the slope to be 0.16 and 0.11 in Fig. 6a; thus, if the Z grains within the shaded area in Fig. 6b came from a 3 $M_\odot$, 0.3 $Z_\odot$ AGB star and the other Y and Z grains from a 2 $M_\odot$, 0.3 $Z_\odot$ AGB star, the calculated $\Delta^{50}Ti_{48}$ values of the two populations of grains would shift downward by 24% and 16%, respectively. Furthermore, here we assumed that the GCE simply follows a linear trend in Fig. 6a, and, in turn, any deviation from the GCE line in the initial composition of the parent star of a grain would directly result in uncertainties in the the calculated $\Delta^{50}Ti_{48}$ value. Besides, while the Si GCE trend has been tested and confirmed by both presolar silicate and SiC grain data [68], the Ti GCE trend is much less well-understood, and our adopted slope-0.65 line is a crude assumption. Given these uncertainties in the derived $\Delta^{50}Ti_{48}$, the $^{13}$C pocket strengths inferred from $\delta^{88}Sr_{87}$ are more reliable, because both $^{87}$Sr and $^{88}$Sr are significantly overproduced during the AGB phase so that $\delta^{88}Sr_{87}$ is barely affected by the initial Sr isotopic composition, i.e., by the GCE.

[6] In comparison, the 2 $M_\odot$, 1.0 $Z_\odot$ FRANEC Torino AGB model calculations in D1.5 to U1.3 cases, which provide a good match to the MS grain data in Fig. 5, predict that the AGB stellar nucleosynthesis falls along slope-0.24 and -0.18 lines, respectively, in Fig. 6a. Thus, both sets of AGB models support our assumption that AGB nucelsoynthesis follows a trend with a slope of 0.24 in the parent AGB stars of MS grains.

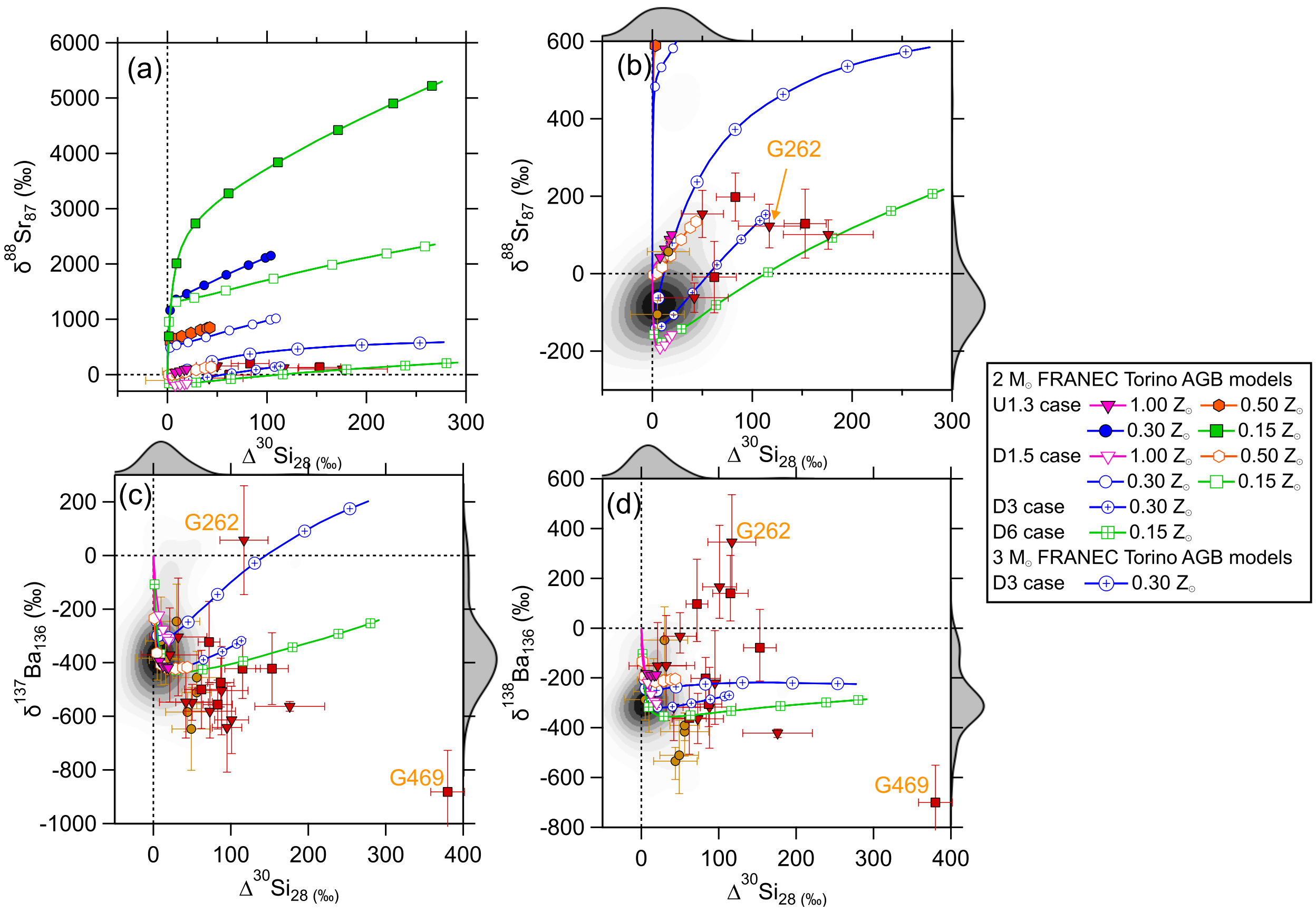

***Figure 5**. Same as Fig. 4 but plotted are FRANEC Torino AGB models. Given the large number of TPs predicted by the FRANEC stellar models, each symbol here represents three TPs. Compared to the reference ST case, the $^{13}$C density is increased by a factor of 1.3 in U1.3 case and reduced by factors of 1.5, 3.0, and 6.0 in D1.5, D3, and D6 cases, respectively. The labels for the models are consistent with those given in [21, 23]. In panels (c) and (d), only plotted are the models that overlap with the grain data in panel (b).*

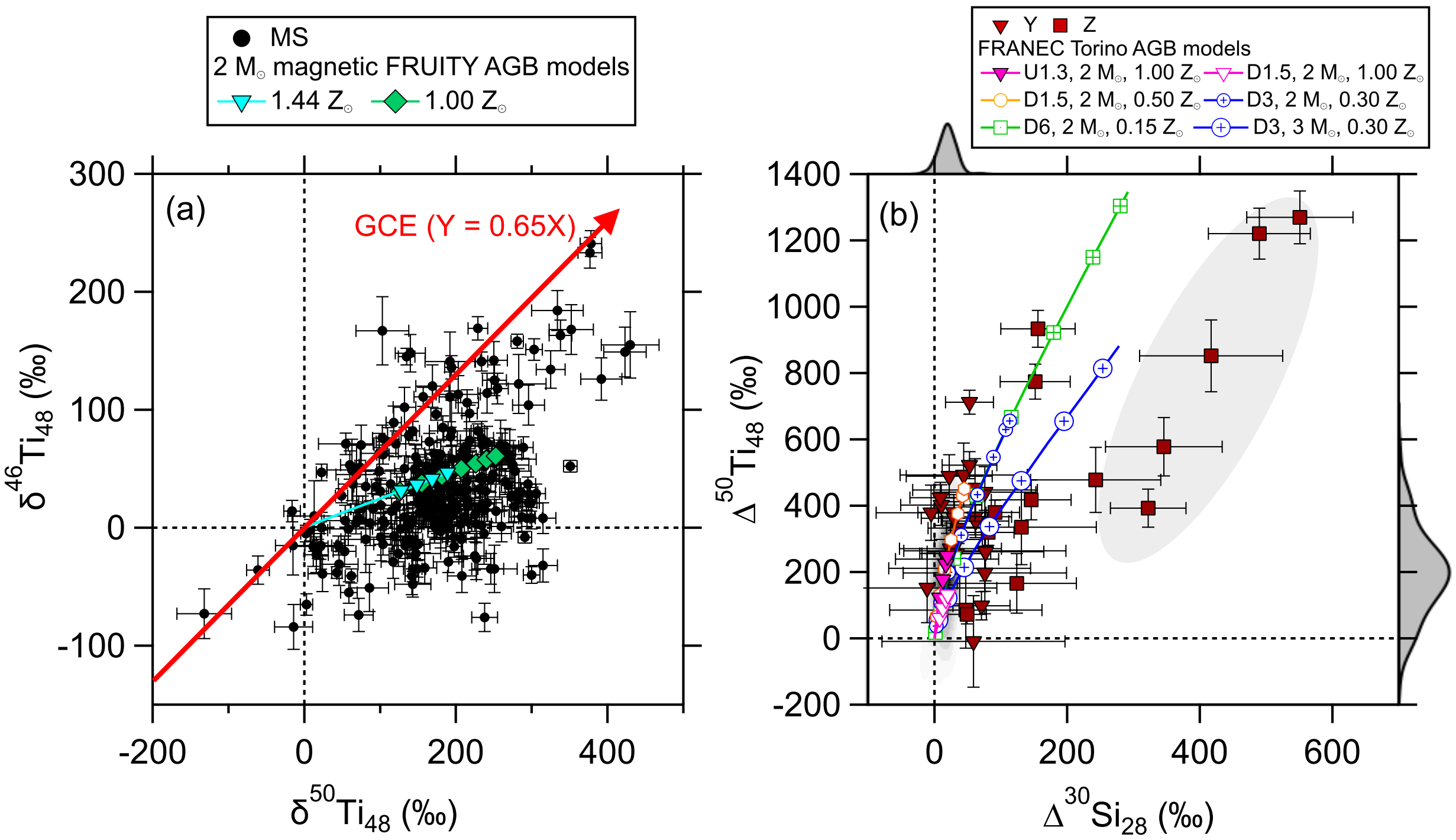

***Figure 6**. In panel (a), we illustrate that like the Si isotope ratios in Fig. 3b the Ti isotope ratios of MS grains receive contributions from both GCE and AGB stellar nucleosynthesis. In panel (b), Y and Z grains are compared to MS grains with 1σ error ($\Delta^{50}$Ti) ≤ 60‰ (greyscale density map) and the FRANEC Torino AGB models that overlap with the Y and Z grains in Fig. 5. The MS, Y, and Z grain data are from [13, 21, 69-74].*

We do not expect that our conclusion here is affected by the adopted $^{13}$C pocket because both the magnetic FRUITY and FRANEC Torino AGB models predict strongly increasing $\delta^{88}$Sr$_{87}$ with decreasing initial stellar metallicity, independent of the adopted $^{13}$C pocket. As pointed out earlier, all existing stellar models, including models calculated using the MESA/NuGrid and Monash codes [48, 75], predict similar trends for the correlation between $\delta^{88}$Sr$_{87}$ and the initial stellar metallicity. The choice of the $^{13}$C pocket, however, affects the correlation between $\delta^{88}$Sr$_{87}$ and $\delta^{138}$Ba$_{136}$ [54] and needs to be investigated based on correlated, higher-precision Sr and Ba isotope data for more Y and Z grains. Indeed, the currently adopted $^{13}$C pocket based on the overshooting mechanism (0.15–0.50 $Z_\odot$ models in the D6 to D1.5 cases) is unable to explain four Y and Z grains with large $\Delta^{30}$Si$_{28}$ and positive $\delta^{138}$Ba$_{136}$ values. In particular, the Y grain M2-A2-G262 exhibits the highest $\delta^{137,138}$Ba$_{136}$ values, pointing to the activation of the $^{136}$Cs ($t_{1/2}$ = 13 d) branch point along the *s*-process and thus an increased operation efficiency of the $^{22}$Ne($\alpha$,$n$)$^{25}$Mg reaction at higher $T_{\max}$. At 0.30 $Z_\odot$, when the initial stellar mass is increased from 2 $M_\odot$ to 3 $M_\odot$, the Torino AGB model in the D3 case can account for the $\delta^{137}$Ba$_{136}$ value observed in M2-A2-G262 but not $\delta^{138}Ba_{136}$, which could imply a unique $^{13}C$ distribution pattern in its parent star (different than that adopted in the Torino AGB models here). In addition to M2-A2-G262, in Fig. 6b Torino AGB models suggest that the Z grains within the shaded area also came from more massive AGB stars than the other Y/Z and MS grains. However, the lowered $\Delta^{50}Ti_{48}$ values of these Z grains at large $\Delta^{30}Si_{28}$ could alternatively be explained by assuming GCE trends with slightly reduced slopes. For instance, if the assumed slope is reduced from 0.65 to 0.50, the highest calculated $\Delta^{50}Ti_{48}$ value for one Z grain changes from 1270‰ to 1840‰, in which case the Z grain would fall close to the trend defined by the 2 $M_{\odot}$ Torino AGB models. Strontium and Ba isotope data are needed to examine whether Z grains with $\Delta^{30}Si_{28} > 200$‰ came from more massive AGB stars, in which case we expect to observe enhanced $\delta^{88}Sr_{87}$ and $\delta^{137}Ba_{136}$ values. In the Z grain M2-A1-G469 with $\Delta^{30}Si_{28} = 380$‰, we, however, observed the opposite – the lowest $\delta^{137,138}Ba_{136}$ values among all the Y and Z grains from this study. Currently, we cannot find a consistent explanation to its Si and Ba isotopic composition using FRANEC Torino AGB models. Higher precision Sr and Ba data are needed for more Y and Z grains to test whether such signatures are common in grains with large $\Delta^{30}Si_{28}$ values.

We note that our constraint ($0.15\ Z_{\odot} \leq Z < 1.00\ Z_{\odot}$) on the metallicity of the parent AGB stars of the Y and Z grains from this study was derived based on 2 $M_{\odot}$ models and is thus affected by uncertainties in the initial stellar mass of their parent stars. With increasing initial stellar mass, we expect to see enhanced $\Delta^{30}Si_{28}$ and $\delta^{88}Sr_{87}$ values (Fig. 5b), which would thus shift our metallicity constraint upward (i.e., higher stellar metallicities); and vice versa. However, based on $\delta^{137}Ba_{136}$ (Fig. 5c) M2-A2-262 seems to be the only grain that came from a more massive AGB star than the rest of the MS, Y, and Z grains. In other words, our grain data do not require systematic differences in the initial parent stellar mass among the three groups of grains.

Our inferred strengths of the $^{13}C$ pocket (D6–D1.5 cases) for the parent low-metallicity AGB stars of Y and Z grains, lie at the lower end of those (D6–U2 cases; D1.5 case on average) inferred for low-metallicity AGB stars based on stellar observations [49]. Both intrinsic AGB and extrinsic *s*-enriched stars[7] exhibit an overall trend of increasing [hs/ls][8] with decreasing metallicity

[7] Intrinsic AGB stars are AGB stars that experience or have experienced *s*-process nucleosynthesis. Extrinsic *s*-enriched stars are stars that have been polluted by a companion AGB star but have not (yet) reached the AGB stage.
[8] ls refers to the abundance of elements at the first *s*-process peak (e.g., Sr), while hs refers to the abundance of elements at the second *s*-process peak (e.g., Ba).

[49, 67], pointing to continuously increasing $^{13}$C/Fe ratio with decreasing [Fe/H] but with some scatter. It was shown in [49] that the overall trend of increasing [hs/ls] with decreasing metallicity can be reproduced by FRANEC Torino AGB model predictions in the D1.5 case for all intrinsic and extrinsic AGB stars and that the scatter in the trend needs to be explained by varying $^{13}$C-pocket strengths (D6–U2 cases). Given that the inferred $^{13}$C-pocket strengths for the parent stars of Y and Z grains are, on average, lower than those observed for AGB stars, it implies that the $^{13}$C-pocket formation efficiencies of the former are not representative of those in present-day low-metallicity AGB stars. The unrepresentativeness of the parent stars of types Y and Z grains may stem from the fact that the grains originated from a limited number of ancient low-metallicity stars, given that Y and Z grains are more than an order of magnitude less abundant than MS grains in primitive meteorites.

The low-metallicity ($0.15\ Z_{\odot} \leq Z < 1.00\ Z_{\odot}$) AGB stellar origins of Y and Z grains, however, are challenged by the indistinguishable Mo isotopic compositions of MS, Y, and Z grains. This is because the FRANEC Torino AGB models predict different Mo isotopic patterns for a 2 $M_{\odot}$, 0.3 $Z_{\odot}$ AGB star than those for a 2 $M_{\odot}$, 1.0 $Z_{\odot}$ AGB star [23], resulting from (*i*) increased $T_{max}$ values in lower-metallicity AGB stars, and (*ii*) the deviation of the Maxwellian-averaged cross sections of $^{95}$Mo, $^{96}$Mo, $^{97}$Mo, and $^{98}$Mo from $1/v_{th}$, in which $v_{th}$ is the thermal velocity. New neutron capture cross section measurements of $^{95,96,97,98}$Mo using state-of-the-art facilities are needed to examine whether the Maxwellian-averaged cross sections of these Mo isotopes indeed deviate from the $1/v_{th}$ rule. Note that although the Mo isotope data for the Y and Z grains from [23] and MS grains from [11] were possibly affected by terrestrial Mo contamination (see discussion in Appendix), Mo contamination would not be able to move MS, Y, and Z grains to the same linear trend in $\delta^{95,97,98}$Mo vs $\delta^{100}$Mo plots. Thus, Mo contamination cannot explain the indistinguishable Mo isotopic compositions of MS, Y, and Z grains (see [23] for details).

## 5. CONCLUSIONS

The Sr and Ba isotope data of the Y and Z SiC grains from this study reveal that Y and Z grains exhibit higher $^{88}$Sr/$^{87}$Sr and more variable $^{138}$Ba/$^{136}$Ba ratios than MS grains. Our comparisons with two sets of AGB stellar nucleosynthesis models suggest that the Si, Sr, and Ba isotopic compositions of the Y and Z grains can be consistently explained only if the grains came from low-metallicity ($0.15\ Z_{\odot} \leq Z < 1.00\ Z_{\odot}$) AGB stars in which the efficiency of the $^{13}$C-pocket formation was greatly reduced. This implication is supported by the $^{49}$Ti and $^{50}$Ti excesses of Y and Z grains [21]. However, it is challenged by the indistinguishable Mo isotopic compositions of MS, Y, and Z grains. Since the varying Mo isotopic patterns predicted by AGB models for stars with different metallicities result mainly from the energy dependences of $^{95,96,97,98}$Mo($n$,$\gamma$) cross sections, new measurements of these neutron capture cross sections across the relevant AGB temperature regime are needed to examine whether their Maxwellian-averaged cross sections deviate from the $1/v_{th}$ rule. If Y and Z grains are confirmed to have originated from low-metallicity AGB stars, their isotopic compositions will provide valuable constraints on poorly known stellar parameters for low-metallicity AGB models.

**Acknowledgements**: We dedicate this paper to the memory of Franz Käppeler, grand master of *s*-process nucleosynthesis and a dear friend to many of the authors. We would like to thank Franz for his lifelong dedication to pushing the limit of neutron capture cross section measurements, without which the scientific return of presolar grain isotope data would be greatly reduced. This work was supported by NASA through grants 80NSSC20K0387 to N.L., NNX10AI63G and NNX17AE28G to L.R.N., and 80NSSC17K0251 and 80NSSC21K0374 to A.M.D. D.V. acknowledges the financial support of the German-Israeli Foundation (GIF No. I-1500-303.7/2019).

## Appendix

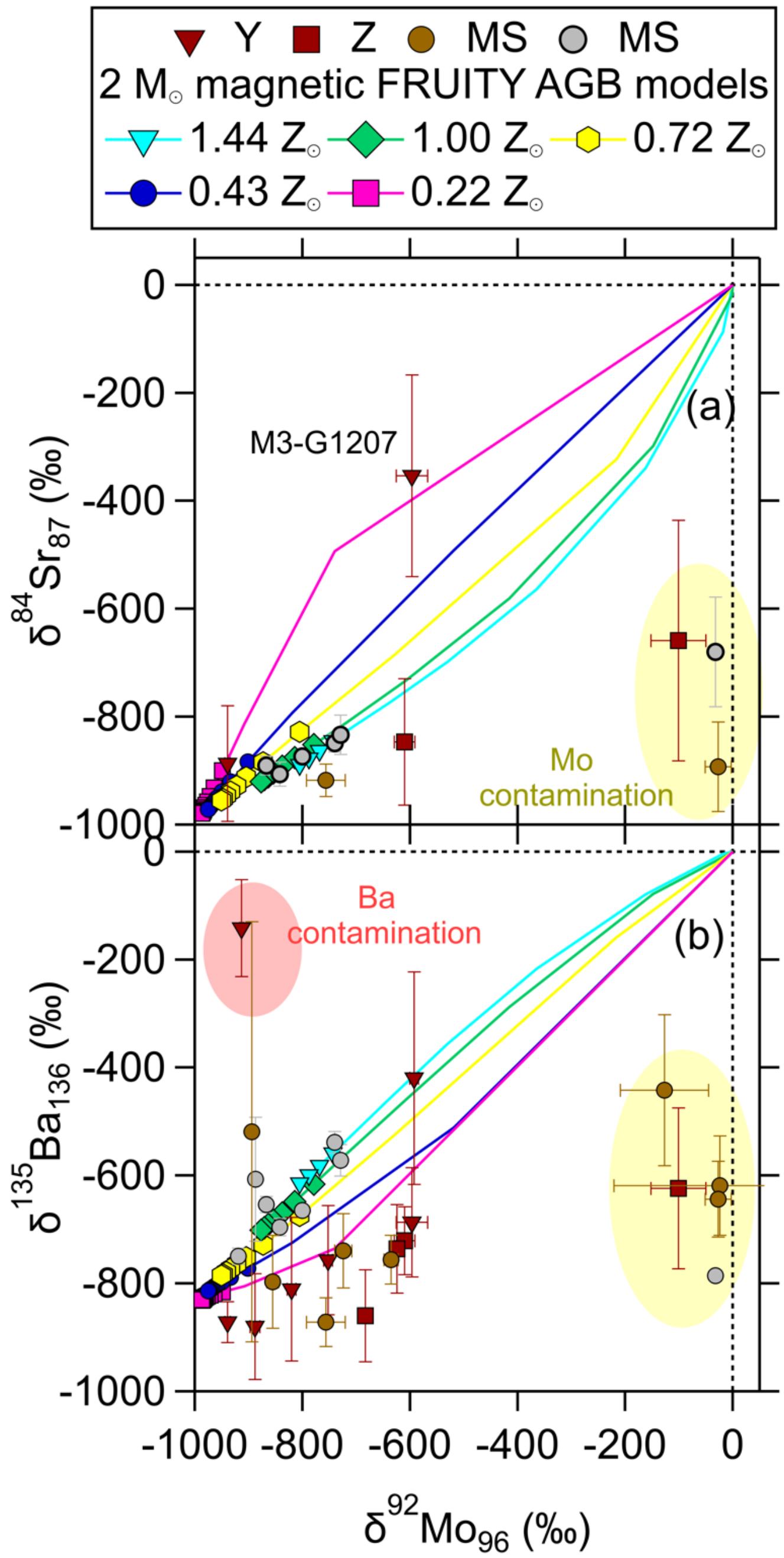

**Fig. A1**. *Plots of Sr-Mo and Ba-Mo isotopes comparing Y and Z grains from this study and MS grains from [11] (brown circles) and [20] (grey circles) with AGB models. The magnetic FRUITY AGB models are the same as those shown in Figs. 3 & 4. Asteroidal/terrestrial Mo and Ba contaminations dominate the Mo and Ba isotopic compositions of grains in yellow and red shaded areas, respectively in panels (a) and (b).*

It was shown that multielement isotope data can be used to identify contaminated grains [56]. Since we obtained the isotope data of more than one heavy element (Sr, Mo, or Ba) for 15 of the 18 Y and Z grains from this study, we chose $\delta^{84}Sr_{87}$, $\delta^{92}Mo_{96}$, and $\delta^{135}Ba_{136}$ for comparing MS, Y, and Z grains from this study and the literature [11, 23] with magnetic FRUITY AGB models in Fig. A1. The three isotope ratios are chosen because they are least affected by uncertainties in AGB model predictions for the *s*-process (see [56] for discussion in detail). Figure A1 reveals a good agreement of the magnetic FRUITY AGB models with all but one MS grain from [20]. In comparison, our Y and Z grains and MS grains from [11, 23], all of which were found on the same sample mounts and analyzed in the same CHILI session, generally lie to the right of the model predictions but agree with the models within 2σ errors. The difference between the MS grains from [20] and the MS/Y/Z grains from [11, 23] could point to varying degrees of Mo contamination, but a definitive conclusion is hampered by the large errors and model uncertainties. As pointed out by [23], the Y, Z, and MS grains from [11, 23] likely sampled some Mo contamination because these grains (0.5−2 μm in size) are smaller than the MS grains (1.5−3 μm) from [20] and are comparable to the laser beam (~1 μm) used for sputtering material in the CHILI instrument [52]. The three Mo-contaminated MS grains from [11] (within yellow shaded area in Fig. A1b) were already noted in that study based on their multielement isotope data.

Except for Z grain M3-692 whose Mo isotope data mainly reflect asteroidal/terrestrial Mo, the Mo isotopic signatures of all the other Y and Z grains are dominated by AGB *s*-process Mo isotopic signatures. We cannot accurately estimate the percentage of asteroidal/terrestrial Mo contamination for each of the grains due to the large errors of their $\delta^{135}Ba_{136}$ values and potential modeling uncertainties. We identified one Ba-contaminated Y grain, M2-A2-G1140 (within red shaded area in Fig. A1b; excluded in Figs. 2−5). Besides, although Y grain M3-G1207 cannot be explained by the magnetic FRUITY models during the C-rich phase, this grain overlaps with the models during the O-rich phase within 1σ errors in both panels of Fig. A1. Since it is highly unlikely that a meteoritic/terrestrial contaminant has Sr/Mo and Ba/Mo ratios that are similar to those of M3-G1207, the isotopic signature of the grain likely implies reduced *s*-process isotope enrichments ($^{87}Sr$, $^{96}Mo$, $^{136}Ba$) in the envelope of its parent AGB star compared to the AGB model predictions.